\documentclass{article}
\usepackage{amsfonts, amssymb, amsmath, amsthm, mathrsfs, chet, cite}
\usepackage{graphicx} 
\usepackage{braket}
\usepackage{bm}
\usepackage{cancel}

\usepackage{subcaption}  
\usepackage{algorithm} 
\usepackage{algpseudocode}

\begin{document}

\title{Bit-bit encoding, optimizer-free 
training and sub-net initialization: techniques for scalable quantum machine learning}
\author{Sonika Johri \email{sjohri@coherentcomputing.com}}
\affiliation{Coherent Computing Inc, Cupertino CA}
\date{}

\abstract{
Quantum machine learning for classical data is currently perceived to have a scalability problem due to (i) a bottleneck at the point of loading data into quantum states, (ii) the lack of clarity around good optimization strategies, and (iii) barren plateaus that occur when the model parameters are randomly initialized. In this work, we propose techniques to address all of these issues. First, we present a quantum classifier that encodes both the input and the output as binary strings which results in a model that has no restrictions on expressivity over the encoded data but requires fast classical compression of typical high-dimensional datasets to only the most predictive degrees of freedom. Second, we show that if one parameter is updated at a time, quantum models can be trained \textit{without using a classical optimizer} in a way that guarantees convergence to a local minimum, something not possible for classical deep learning models. Third, we propose a parameter initialization strategy called sub-net initialization to avoid barren plateaus where smaller models, trained on more compactly encoded data with fewer qubits, are used to initialize models that utilize more qubits. Along with theoretical arguments on efficacy, we demonstrate the combined performance of these methods on subsets of the MNIST dataset for models with an all-to-all connected architecture that use up to 16 qubits in simulation. This allows us to conclude that the loss function consistently decreases as the capability of the model, measured by the number of parameters and qubits, increases, and this behavior is maintained for datasets of varying complexity. Together, these techniques offer a coherent framework for scalable quantum machine learning.
}

\maketitle

\section{Introduction}\label{sec:intro}

The most practical and generally applicable formulation of quantum machine learning at the current time relies on variational quantum circuits with free parameters that can be trained to model the correlations in a dataset analogous to classical neural networks \cite{Wang_2024}. The current approach to training a predictive model in this paradigm requires data samples to be loaded into the quantum computer one at a time. After the execution of a parametrized quantum circuit, the qubits are measured, and the outcomes are classically processed to calculate the value of a loss function. This value is then submitted to a classical optimizer which decides the next set of parameter values for which the circuit is evaluated on the quantum computer. Training a generative model proceeds similarly with the difference that the data is read out at the end instead of being loaded. Quantum machine learning models have been tested on near-term quantum hardware for various real-world datasets \cite{Gujju24}, \cite{elton_copula}, \cite{zhu2022copulabased}, \cite{iaconis2023tensor}, \cite{Johri20}, \cite{Rudolph20}, \cite{cherrat2022quantum}, \cite{silver2023mosaiq}, \cite{Silver_Patel_Tiwari_2022}.

Although quantum models can typically surpass the  simulation capabilities of classical computers, provable arguments for quantum advantage often only apply in restricted contexts \cite{Liu2021}, \cite{Lloyd2016}. Since real world datasets do not neatly fit into a computational complexity class, a clear picture of the efficacy of quantum vs. classical models can really only be obtained by training them on these datasets and studying their performance as a function of scaling the number of qubits and number of gates for different architectures. In this work, we therefore focus on techniques that enable the training of quantum machine learning models at scale.

For any family of machine learning models, a principal criteria for their usefulness is that a form of universal approximation exists with respect to the function that describes the target dataset. In quantum machine learning, the functional form of the model depends on the data encoding. Generally speaking, amplitude encoded data is `dense', that is, the amount of encoded information is linear in the size of the Hilbert space in which the model operates, and the model is linear in the encoded data. On the other hand, angle encoded data is `sparse', that is, the amount of encoded information is exponentially small in the size of the Hilbert space but the model can be more expressive. \cite{Schuld_data_encoding} showed that quantum models can be analyzed as a partial Fourier series in angle encoded data, with coefficients determined by the parameterized circuit, and frequencies by the eigenvalues of the encoding operators. In either case, the time for loading is lower-bounded by the quantity of information to be loaded. Even an image from a simple dataset like MNIST which has 1024 pixels cannot in the near-term be exactly loaded into a quantum state. Since this limit is fundamental, it is clear that some kind of compression is required before loading. Previous work has shown the effectiveness of using tensor network techniques to compress the data before loading it as an amplitude encoded state \cite{iaconis2023quantum}, \cite{iaconis2023tensor}, \cite{ pollmann_tensor}, \cite{jumade2023dataloadableshortdepth}. Other common strategies involve using techniques like principal component analysis (PCA) or variational autoencoders to compress the data by reducing its dimensionality. 

Here, we propose the use of an ultra-sparse encoding scheme in which both the input and output of the quantum model are bit strings, which we call `bit-bit' encoding. The universal approximation for this encoding holds for \textit{any} function between the input and output bits. To make this encoding effective, we propose an efficient classical binary encoding scheme that extracts the most predictive bits to represent samples from the real-valued dataset. Similar schemes are often used in classical data applications. For example, a scheme called iterative quantization is used for binary encoding of images for the purpose of large-scale image retrieval \cite{itq}. In quantum machine learning, the bit (or basis) encoding has previously been used for generative learning of joint probability distributions via qopula states \cite{elton_copula}, \cite{zhu2022copulabased}.

Another essential criteria for a machine learning paradigm is the ability to train the model to convergence. For classical neural networks, gradient descent is practically found to work well. Essential to this is the ability to pick good hyperparameters such as the learning rate. This is usually done through trial-and-error. A learning rate that is too high can lead to oscillations or overshoot the minimum. A learning rate that is too small can take unnecessarily long to converge. Often the learning rate is chosen to have a decaying or oscillating schedule rather than a constant value. In all cases, there is no guarantee that the chosen learning rate will lead to the model converging to a local minimum. On quantum computers, tuning of learning rates using automated Bayesian optimization or genetic algorithms can be especially expensive  \cite{hpo}. Alternatives to full gradient descent for training variational quantum circuits such as variants of SPSA \cite{Wiedmann_2023} and coordinate descent \cite{random_coordinate_descent} face similar challenges in setting hyperparameters.

In contrast to the above techniques, here we show that it is possible to train a variational quantum circuit without using an optimizer at all! This is achieved via `exact coordinate updates' - updating the model one parameter at a time using an analytic expression for its minimum that takes only 2 measurements upto a desired precision per data sample as arguments, which is the same sample complexity as gradient descent. We show that while gradient based methods would need to have a decaying learning rate as the training progresses, and get stuck at saddle points, our method offers guaranteed convergence to a local minimum. In this way, it confers a significant advantage to large quantum models even compared to classical deep learning for which such a guarantee is not available. 

Yet another criteria that has emerged as an obstacle for large quantum machine learning models involves the loss function landscape on random initialization becoming exponentially flat as the number of qubits increases. This, in turn implies that an exponential number of measurements is required to determine the parameter updates. This phenomenon is known as a barren plateau \cite{McClean2018}, and affects both gradient-based and gradient-free optimizers \cite{Arrasmith2021effectofbarren}. The presence of barren plateaus has been linked to expressivity of the ansatz, and consequently, some strategies to avoid them involve constraints on the structure of the quantum ansatz which, however, can lead to sub-optimal results \cite{PRXQuantum.3.010313} or even dequantization \cite{gilfuster2024relationtrainabilitydequantization}. Other strategies include smart initialization \cite{verdon2019learninglearnquantumneural} which requires use of classical neural networks, layerwise training \cite{Skolik2021}, and initialization of part of the model as an identity operator \cite{Grant2019initialization}.

While the above methods of avoiding barren plateaus work for fixed input and output data, here we propose an iterative approach which starts off with a small model trained on lower-dimensional (more compressed) input data. At each subsequent stage, the current model is used to initialize corresponding parameters of a larger model which acts on higher-dimensional (less compressed) data and has more parameters. The operators in the larger model that are not in the sub-net are initialized as identity. This technique, which we call sub-net initialization, allows previously trained quantum models to be used for training larger models as more quantum computational resources become available.

We numerically test these methods on the MNIST dataset with systems of 4-16 qubits on quantum models that have full connectivity between qubits. The emphasis of our testing is to validate scalability as the size of the model, quantified by the number of the qubits and number of parameters, and the hardness of the learning problem, quantified by the number of classes increases. This is in contrast to many studies which emphasize stand-alone performance of a small quantum model, often coupled to a large classical neural network in a hybrid architecture, without presenting an understanding of whether there are performance benefits as the quantum components of the architecture are scaled up. The latter is necessary to understand at what point we may expect machine learning on quantum systems to become competitive with classical.

The outline of this paper is as follows. Sections \ref{sec:bit-bit}, \ref{sec:coord_update} and \ref{sec:subnet} introduce the bit-bit encoding, exact coordinate update training, and sub-net initialization respectively. Section \ref{sec:numerics} presents numerical results from training the models on a quantum simulator using these techniques. Section \ref{sec:discussion} discusses the results and suggests directions for future work.

\section{Bit-Bit Encoding}\label{sec:bit-bit}
\subsection{Formulation}
Consider a supervised learning problem in which we are given a dataset $\mathcal{D}=\{x,y\}$ which consists of data points $x=(x_1, x_2, \ldots)\in\mathcal{X}$ with accompanying labels $y\in\mathcal{Y}$, with the number of classes $|\mathcal{Y}|=N_c$. Without loss of generality, we assume that $\mathcal{X}\in[0,1]^d$,  and $\mathcal{Y}\in\{0, 1, \ldots, N_c-1\}$. We set $N_{\text{data}}$ as the number of available data samples. For a classification problem, the learning task consists of finding an efficiently computable function that maps $\mathcal{X}$ to $\mathcal{Y}$. Typically, the function is many-to-one, that is, $N_c<|\mathcal{X}|$.

Now let's find a binary encoding that approximates the original dataset as $\mathcal{D}_b=\{z_b,y\}$, where $b$ is a vector of natural numbers of dimension $d$ with $\sum_i b_i=B$. $z_{i}$ is the value of $x_i$ truncated to $b_i$ bits of precision, that is, $z_{ib} = \left\lfloor x_i 2^{b_i} \right\rfloor$. We include the case when $b_i=0$ for some components $i$. Henceforth, the argument $b$ will be dropped from $z$. As the magnitude of the $b_i$ grow, $\mathcal{D}_{b}$ more precisely captures the original dataset.

In this compressed dataset, there may be collisions at the input, so that a particular value of $z$ may occur more than once, each time corresponding to the same or different value of $y$. Then, the available data can be considered as samples from a joint probability distribution $f(\mathcal{Z}=z, \mathcal{Y}=y)$ over correlated random variables $(\mathcal{Z}, \mathcal{Y})$. For the purpose of classification, the `correct' output is then taken to be the mapping which occurs more frequently, that is 
\begin{align}
    C(z) = \arg \max_y f(z,y),
\end{align}
is the classification function to be learned.

We see that $z$ can be exactly loaded into a quantum state that lives in the Hilbert space of $N_x \geq B$ qubits using Pauli $X$ gates acting on the $\ket{0}$ state to create the computational basis state $\ket{z}$. As $N_x$ grows, the representation of the dataset in the Hilbert space of the qubits becomes more exact. Similarly, the labels can be exactly mapped onto computational basis states $\ket{y}$ of $N_y=\left\lceil \log_2{N_c} \right\rceil$ qubits. Thus the classification function is of the form $C: \{0, 1\}^{N_x} \to \{0, 1\}^{N_y}$. We have thus approximated the problem of learning a multivariate function of real numbers to that of learning a vectorial Boolean function. Since both the input and output are now represented as binary strings, we call this a bit-bit encoding.

The quantum learning problem can now be cast in terms of a unitary $U_*$ that has the following action on $N_q=N_x+N_y$ qubits:
\begin{equation}\label{eq:bit_bit_unitary}
    \ket{0}\ket{z}\xrightarrow{U_*} \ket{C(z)}\ket{g(z)}
\end{equation}
where the states $\ket{g(z)}$ and $\ket{z}$ have support on $N_x$ qubits, and $N_y$ qubits initialized in state $\ket{0}$ are mapped by the unitary to the state $C(z)$. For the purpose of classification, we can discard the $g$ states. However, note that if $C(z)=C(z')$, we have $\langle g(z) | g(z')\rangle =\delta_{z,z'}$.

Thus, we can formulate the learning task on the quantum computer as that of finding a unitary $U$, parameterized by variables $\theta$, such that when it acts on $\ket{0}\ket{z}$, the probability of measuring $\ket{C(z)}$ at the output is maximized. That is, given,
\begin{align}
    U(\vec{\theta})\ket{0}\ket{z}=\sum_k \sqrt{P_{k,z}(\vec{\theta})}e^{i\phi_{k,z}(\vec{\theta})}\ket{k}\ket{g_{k,z}(\vec{\theta})},
\end{align}
find the value of $\vec{\theta}$ which minimizes the loss function
\begin{align}\label{eq:loss}
    \bar{L}(\vec{\theta})=1-\sum_{z\in \mathcal{D}_{\vec{b}}} f(z) P_{C(z),z}(\vec{\theta}),
\end{align}
where $f(z)$ is the frequency of occurence of $z$ in the dataset. $0\leq \bar{L}\leq 1$, and has the simple interpretation of being the probability of the model giving the wrong answer when queried. If $U(\vec{\theta})=U_*$ and there is no noise, even one shot would suffice to classify the input data sample.

Note that the loss function only includes the action of $U$ on values of $z$ in the dataset. The action of $U$ on values of $z$ not in the dataset will determine its generalization behavior. In that case, a quantum model that is perfectly trained on the dataset will output a probability distribution over the possible classes. 

\subsection{Expressivity}
If there are no restrictions on the number of quantum operations to implement $U(\vec{\theta})$, it can get arbitrarily close to $U_*$, approximating any function between the input and output. Therefore, in the limit of infinite gates, the quantum circuit has perfect expressivity. This is in contrast to other encodings such as amplitude or angle encoding, where the class of functions that the model can learn is restricted by the encoding method no matter how many gates are in the circuit.

That this ultra-sparse encoding saturates the expressive capabilities of a quantum model reflects a general trend in quantum machine learning, namely that the denser the encoding, the more restricted the class of functions the circuit can approximate. Table \ref{tab:expressivity} show this relationship for three common encoding types. We define information density as the information capacity of the data loading steps divided by the size of the Hilbert space in which the model operates:
\begin{align}
    \kappa_{\text{info}} = \frac{\text{Information capacity of data loading operations}}{\text{Size of Hilbert space accessible to model}}
\end{align}

\begin{table}[h]
    \centering
    \begin{tabular}{|c|c|c|}
    \hline
        Encoding Type & Expressivity & $ \kappa_{\text{info}}$\\
    \hline
        Amplitude & Linear & $O(2^m)$\\
    \hline
        Angle & Trigonometric Polynomial & $O(N_q 2^m N_H^{-1})$\\
    \hline
        Bit-Bit & Full & $O(N_q N_H^{-1})$\\
    \hline
    \end{tabular}
    \caption{Relationship between expressivity and information density of the data loading steps of the quantum learning algorithm in the limit of infinite circuit depth. Here $m$ is the number of bits of precision when a real number is being loaded in the amplitude or angle encoding methods. $N_H$ is the size of the Hilbert space accessible to the model. For models with large number of qubits, typically $N_H>>2^m$ and $N_H>>N_q$.}
    \label{tab:expressivity}
\end{table}

Note that some models may not operate in the entire Hilbert space of the qubits such as the quantum nearest centroid algorithm demonstrated in \cite{Johri20}. This quantum nearest centroid algorithm is also an example of quantum advantage from amplitude encoding despite it only having linear expressivity.

\subsection{Binary Encoding}
In classical data applications where computational speed and storage efficiency are critical, such as large-scale data analysis or retrieval systems, an encoding scheme is used to represent high-dimensional data with binary vectors. Techniques for binary encoding like iterative quantization \cite{itq} and spectral hashing \cite{spectral_hashing} have shown good performance for such applications.

Here we have a similar problem in that we would like to find a compact but predictive binary encoding of the dataset, which also scales efficiently with the number of dimensions of a data sample. We thus design a technique based on extracting the most predictive bits. Our binary encoding scheme proceeds as shown in Algorithm \ref{alg:binary_encoding}. The goal is to encode each data sample into a bitstring of a given length. We first use PCA to compress the original data to a smaller number of dimensions $D\leq d$, where $d$ is the number of original dimensions of each data sample. Then for each PCA direction, we calculate an importance score based on the mutual information between the data distribution along that direction and the output. Finally, we assign a number of bits to each dimension according to the ratio of its importance score and the sum of all the importance scores. Thus, we get an encoding that ensures that more `important' directions, that is, those that contain more predictive information, are assigned more bits.

\begin{algorithm}
\caption{Mutual Information-Based Binary Encoding Scheme}
\begin{algorithmic}[1]
\Require $X$: Input data, $Y$: Output labels, $B$: Total number of bits
\Ensure Bitstrings for each data sample

\State \textbf{Step 1: Dimensionality Reduction}
\State Perform PCA on $X$ to reduce it to $D$ dimensions: $X \gets \text{PCA}(X, D)$

\State \textbf{Step 2: Calculate Importance Scores}
\For{each PCA direction $i = 1$ to $D$}
    \State Compute importance score $I_i$ as the mutual information between $X_i$ and $Y$
\EndFor

\State \textbf{Step 3: Assign Bits to Dimensions}
\For{each dimension $i = 1$ to $D$}
    \State Assign bits $b_i = \lfloor B \cdot \frac{I_i}{\sum_{i'=1}^D I_{i'}} \rceil$
\EndFor

\State \textbf{Step 4: Encode Data Samples}
\State Encode each data sample in $X$ using the assigned bit allocation $\{b_1, b_2, \dots, b_D\}$
\end{algorithmic}\label{alg:binary_encoding}
\end{algorithm}

The scheme ensures that as more qubits become available, the new encoding encapsulates the smaller encodings, which will become important in the sub-net initialization scheme presented in Section \ref{sec:subnet}. Table \ref{tab:encoding_4_classes} shows how Algorithm \ref{alg:binary_encoding} generates a distribution of bits over the most relevant PCA directions for the classes 0, 1, 2, and 3 of the MNIST dataset. 

\begin{table}[H]
    \centering
    \begin{tabular}{|c|l|c|}
    \hline
        $B$&Bit Distribution & $p$  \\
        \hline
          2&[1,1]&69\\
          4&[1,1,1,1]&204\\
          6&[2,1,1,1,1]&321\\
          8&[2,1,1,1,1,1,1]&726\\
          10&[3,1,1,1,1,1,1,1]&915\\
          12&[3,2,1,1,1,1,1,1,1]&1140\\
          14&[4,2,2,1,1,1,1,1,1]&1401\\
         \hline              
    \end{tabular}
    \caption{The distribution of $B$ bits among the PCA directions that explain the most variance (in order from left to right) and the number of parameters $p$ in the model for the 4 class training discussed in section \ref{sec:numerics}. Note that smaller encodings are subsets of larger ones. In this class of models we take $N_y=2$ qubits to encode the 4 classes as bit strings and the total number of qubits $N_q = B+2$.}
    \label{tab:encoding_4_classes}
\end{table}

The time for this binary encoding scheme is linear in the number of dimensions as seen from the loops in Algorithm \ref{alg:binary_encoding} while the time for calculating the mutual information along each direction is primarily set by the time to cluster the data. The binary encoding time then scales as $\mathcal{O}(D N^2_{\text{data}})$. Overall, including the time to perform PCA, the classical pre-processing time scales as $\mathcal{O}(d^2 N_{\text{data}} +d^3+D N_{\text{data}}^2)$.

\subsection{Quantum Advantage}
With this encoding, quantum advantage for a particular dataset will exist when $U_*$ can be approximated efficiently on a quantum computer but not classically. That is, with the quantum model, we aim to learn a function that is $\mathcal{O}(\exp(N_q))$ to compute classically while being $\mathcal{O}(\text{poly}(N_q))$ on a quantum computer. Eq. \ref{eq:bit_bit_unitary} encompasses unitaries based on Boolean functions which are the basis of many quantum algorithms with proven exponential speed-up such as Shor's algorithm for factoring. In fact, one of the few quantum machine learning algorithms with proven exponential speed-up, which is based on an artificial dataset constructed from a discrete-log problem \cite{Liu2021}, also fits into this framework. Most datasets however will not fit neatly into a compactly defined function, and often the only way to determine whether there is quantum advantage will be through actually training quantum models. 

We note that since many Boolean functions can be implemented more efficiently, sometimes exponentially so, when ancilla bits are available for computation \cite{saeedi_reversible}, it seems likely that further reducing $\kappa_{\text{info}}$ so that the number of qubits available for the model is much larger than the number of input bits, that is, $N_q>>B$, will enhance the potential for quantum advantage. We leave the investigation of this to future work.

\section{Model Training via Exact Coordinate Updates}\label{sec:coord_update}
\subsection{Formulation}
Currently, optimization of quantum machine learning models mostly depends upon techniques that are also used to train classical neural networks or optimization problems. The main workhorse here is gradient descent which proceeds as follows: (i) Initialize the parameters randomly or based on some rule. (ii) Calculate the loss function at these parameters. (iii) Calculate the partial gradients of the loss function with respect to the components of the parameter vector. (iv) Change the parameters in the opposite direction of the gradient scaled by a constant known as the learning rate. Variants of vanilla gradient descent include the ADAM optimizer which includes historical information and weights the learning rate for each parameter by the gradient history, and more stochastic versions such as the simultaneous perturbation stochastic approximation (SPSA) algorithm. Recently coordinate descent, where one coordinate is updated at a time by an optimizer according to a gradient calculation, has also been studied for quantum machine learning. Ref \cite{random_coordinate_descent} shows that the number of model evaluations of the random coordinate descent optimizer is no worse than the gradient descent optimizer which updates all coordinates at the same time.

All of these techniques rely on using a classical optimizer to propose parameters which are then used in the cost function evaluation on the quantum computer. All of these require setting various optimizer hyperparameters, including at the very least, a learning rate. \cite{Moussa2024} finds that the learning rate is the most important hyperparameter for determining performance of quantum machine learning models, even more so than the choice of entangling gates, batch size, depth, and use of data reuploading, amongst other tested hyperparameters. For both classical and quantum learning, theoretically, the learning rate for gradient descent type algorithms follows from Lipschitz continuity conditions. Practically, it is set by trial-and-error, and often a learning rate schedule varies it over the course of the training.

Remarkably, it is possible to train a variational quantum model without relying on an any kind of external optimizer at all, which means that the learning rate and other optimizer hyperparameters do not need to be set. A similar result has been presented in papers focusing on chemistry applications \cite{vidal2018calculusparameterizedquantumcircuits}, \cite{PhysRevResearch.2.043158}, \cite{parrish2019jacobidiagonalizationandersonacceleration}, \cite{PhysRevResearch.3.033083} which we derive here for the quantum machine learning loss function in Eq. \ref{eq:loss}. Further, we show that while gradient based optimizers would need to adapt a decaying learning rate schedule, which means the model improves more slowly as the training progresses, the technique presented here encounters no such slowdown.

Consider an ansatz in which each parameter $\theta_j$ in the ansatz corresponds to a unitary $e^{i\theta_j P_j}$, where $P_j$ is a tensor product of operators chosen from the Pauli operators $\{X, Y, Z\}$. Let the ansatz consist of $p$ such parameterized operators interspersed with fixed unitary operators $V_i$.

Then,
\begin{align}\label{eq:U_exp_paulis}
    U= \prod_{j=1}^p U_j =V_pU_p\ldots U_3 V_3 U_2 V_1 U_1 = \prod_{j=1}^p V_j\bigg(\cos\bigg(\frac{\theta_j}{2}\bigg) - i \sin\bigg(\frac{\theta_j}{2}\bigg) P_j\bigg).
\end{align}

Here, we will use the notation `'$:j$ and `$j:$' in the subscript to refer to the matrix product $V_{j-1}U_{j-1} \ldots U_2 V_1 U_1$ and  $V_pU_p\ldots U_{j+2} V_{j+1}U_{j+1}V_j$ respectively. Note that this excludes ansatzes in which parameters for different Pauli rotations are correlated. Other than this restriction, the form of the ansatz above is typical of those used for variational algorithms.

Again without loss of generality, let's say the loss function can be written as a weighted sum over expectation value of Hermitian operators $\hat{O}(z)$ measured on a starting state $\ket{\Psi_0(z)}$ acted upon by $U(\vec{\theta})$,
\begin{align}
    \bar{L}(\vec{\theta}) = \sum_z f(z)\bra{\Psi_0(z)} U^{\dagger}(\vec{\theta}) \hat{O}(z) U(\vec{\theta}) \ket{\Psi_0(z)}
\end{align}

If we freeze all other parameters other than the $j$-th one,
\begin{align}
    \bar{L}(\theta_j) = \sum_z f(z)\bra{\Psi_0(z)}U^{\dagger}_{:j-1} \bigg(\cos\bigg(\frac{\theta_j}{2}\bigg) + i \sin\bigg(\frac{\theta_j}{2}\bigg) P_j\bigg) U^{\dagger}_{j+1:}\hat{O}(z)U_{j+1:} \bigg(\cos\bigg(\frac{\theta_j}{2}\bigg) - i \sin\bigg(\frac{\theta_j}{2}\bigg) P_j\bigg)U_{:j-1}\ket{\Psi_0(z)}
\end{align}

This expression can be simplified to obtain
\begin{align}\label{eq:l_single}
    \bar{L}(\theta_j) = 1-\alpha_j -\gamma_j \cos(\theta_j) -\sigma_j \sin(\theta_j),
\end{align}
where the constants $\alpha_j$, $\gamma_j$ and $\sigma_j$ are independent of $\theta_j$.

We see that this can be rewritten as 
\begin{align}\label{eq:loss_j}
    \bar{L}(\theta_j) = 1 - \alpha_j - \sqrt{\gamma_j^2 +\sigma_j ^2} \cos(\theta_j - \theta_j^*),
\end{align}
where $\theta_j^*$ is a function of $\gamma_j$ and $\sigma_j$. Thus, the loss function will be minimized when $\theta_j=\theta_j^*$. We can find $\theta_j^*$ by evaluating the loss function at angle increments of $\pi$ and $\pi/2$ which give:
\begin{align}\label{eq:L values}
    \bar{L}(\theta_j+\pi) = 1 - \alpha_j + \sqrt{\gamma_j^2 +\sigma_j ^2} \cos(\theta_j-\theta_j^*)\nonumber\\
    \bar{L}(\theta_j+\pi/2) = 1 - \alpha_j +\sqrt{\gamma_j^2 +\sigma_j ^2} \sin(\theta_j-\theta_j^*)
\end{align}

From this, we see that 
\begin{align}\label{eq:update}
\theta_j^*=\theta_j-\tan^{-1}\bigg(\frac{2\bar{L}(\theta_j+\pi/2)-\bar{L}(\theta_j+\pi)-\bar{L}(\theta_j)}{\bar{L}(\theta_j+\pi)-\bar{L}(\theta_j)}\bigg),
\end{align}
appropriately assigning the quadrant by using $\text{sgn}(\cos(\theta-\theta_j^*))=\text{sgn}(\bar{L}(\theta_j+\pi)-\bar{L}(\theta_j))$.

Thus, the sub-problem of finding the optimal value of $\theta_j$ can be solved exactly without a call to a classical optimizer. After updating $\theta_j$, one can move on to a different parameter $j'\neq j$ and update that, and so on. The parameter to be updated can be chosen randomly, sequentially, or based on some other rule. This parameter-by-parameter update is similar to what happens in coordinate descent, however this is not a gradient-based method. Unlike in gradient or coordinate descent, the cost function is guaranteed to decrease at each step because we can change the parameter exactly to reach the minimum possible loss. Since no classical optimizer is required, there is no need to set the learning rate. 

In order to gain a physical understanding of the training dynamics, we can further examine the expressions for $\alpha_j$, $\gamma_j$ and $\sigma_j$. The initial states and observables in the bit-bit encoding are:
\begin{align}
    \ket{\Psi_0(z)}=\ket{z}\nonumber\\
    \hat{O}(z) =1- \ket{C(z)}\bra{C(z)}
\end{align}

Let's define the following quantities which measure the effect of removing or including the full Pauli at the $j$-th gate:
\begin{align}
    a_{j}(z) = \langle C(z)|U_{j+1:} U_{:j-1}|z\rangle\nonumber\\
    b_{j}(z) = i\langle C(z)|U_{j+1:} P_j U_{:j-1}|z\rangle
\end{align}

Based on these definitions, we have the following expressions
\begin{align}
    \alpha_j = \sum_z f(z)\frac{ |a_{j}(z)|^2 + |b_{j}(z)|^2}{2}\nonumber\\ 
    \gamma_j = \sum_z f(z)\frac{ |a_{j}(z)|^2 - |b_{j}(z)|^2}{2}\nonumber\\ 
    \sigma_j = \sum_z f(z)\text{Re}(a_{j}(z)b^*_{j}(z)).
\end{align}
Thus $\gamma_j$ and $\sigma_j$ are measures of how much the cost function changes when the Pauli operator is applied. The maximum change in the cost function when parameter $j$ is updated is $2\sqrt{\gamma_j^2 +\sigma_j ^2}$ which is bounded between 0 and 1. Note that we do not actually have to explicitly calculate $\gamma_j$ and $\sigma_j$ during the training. Instead they are implicit in the coordinate update in Equation \ref{eq:update}. However, a fast way to estimate which coordinates are likely to have large $\gamma_j$ and $\sigma_j$ has the potential to speed up the training, and will be explored in future work.

Finally, we observe that the formula in Eq. \ref{eq:update} will work for any variational cost function that is a linear sum of expectation values. In fact, this technique will work for any variational quantum algorithm parameterized by independent Pauli rotations, such as those used in chemistry or optimization, and is not restricted to quantum machine learning. For cost functions that are non-linear, a more complicated equation will need to be solved to find the optimal coordinate.

\subsection{Advantage over gradient methods}
We now analyze how the above technique has an advantage over other optimization methods, especially gradient based ones. For quantum machine learning, the sample complexity scales with the number of parameters for gradient descent, and this is also true for the method presented above. The essence of its advantage instead lies in the fact that the learning rate does not need to be specified. To understand this, let us look at how the learning rate is theoretically determined from Lipschitz continuity conditions.

We say that the derivative of a multivariate function is overall $K$-smooth in the sense of Lipschitz continuity when
\begin{align}
    ||\nabla \bar{L}(\vec{\theta})- \nabla \bar{L}(\vec{\theta}')||\leq K ||\vec{\theta}-\vec{\theta}'||,
\end{align}
where $||.||$ is the $l^2$-norm.

For gradient descent, at each step the parameter update is given by
\begin{align}
    \vec{\theta}(t+1)=\vec{\theta}(t) - \eta \nabla \bar{L}(\vec{\theta}(t)).
\end{align}

Here, $\eta$ is the learning rate and should be set as $\approx 1/K$. If we do gradient descent with this learning rate, a Taylor series analysis shows that the change in the cost function at time $t$ upto the second order in the gradient is given by \cite{ml_course}
\begin{align}\label{eq:gd_taylor}
    \Delta \bar{L}(t+1) = \bar{L}(t+1) - \bar{L}(t) \approx -\frac{1}{2K} ||\nabla \bar{L}(t)||^2
\end{align}

Similarly, we can define $K_j$-smoothness along the $j$-th component as 
\begin{align}
    |\nabla \bar{L}(\vec{\theta})- \nabla \bar{L}(\vec{\theta} +\vec{e}_j h)|\leq K_j |h|,
\end{align}
where $\vec{e}_j$ is a unit vector along the $j$-th direction. From Eq. \ref{eq:loss_j}, we see that here we should set $K_j=\sqrt{\gamma_j^2+\sigma_j^2}$. 

Using Eq. \ref{eq:loss_j} and \ref{eq:gd_taylor}, the reduction per number of cost function evaluations, $N_{\text{eval}}$, for gradient descent is
\begin{align}\label{eq:gd_update}
    \frac{\Delta \bar{L}_{\text{gradient}}}{N_{\text{eval}}} = -\frac{1}{2K(2p)}|\nabla L|^2 = -\frac{1}{4pK} \sum_{j=1}^p K_j^2 \sin^2(\theta_j -\theta_j^*),
\end{align}
where the brackets in the denominator are the number of loss function evaluations required.

We can similarly analyze coordinate descent, where one parameter is updated at a time,
\begin{align}
    \frac{\Delta \bar{L}_{\text{coordinate}, j}}{N_{\text{eval}}} = -\frac{1}{2K_{\max}(2)}|\nabla L|^2 = - \frac{K_j^2}{4K_{\max}} \sin^2(\theta_j -\theta_j^*),
\end{align}
where, in the absence of an ability to efficiently calculate $K_j$ for each coordinate, $\eta=1/K_{\max}$ is used.

For the exact coordinate update discussed here, the reduction at each step is given by
\begin{align}\label{eq:ec_update}
    \frac{\Delta \bar{L}_j}{N_{\text{eval}}} = - \frac{K_j}{(2)}2\sin^2\bigg(\frac{\theta_j -\theta_j^*}{2}\bigg) = - K_j\sin^2\bigg(\frac{\theta_j -\theta_j^*}{2}\bigg)
\end{align}

We know that $K\geq K_{\max}$, where $K_{\max}=\max\limits_{j}K_j=\max\limits_{j}{\sqrt{\gamma_j^2 +\sigma_j ^2}}$. Thus, since the step size is always constrained by the largest $K_j$, if the $K_j$ are highly anisotropic, gradient-based schemes will be slower than the exact coordinate update method presented here. Therefore, it becomes important to analyze the distribution of the $K_j$ values to compare the different optimization schemes. This can be done numerically as we discuss below.

\begin{figure}
    \centering
    \includegraphics[width=\linewidth]{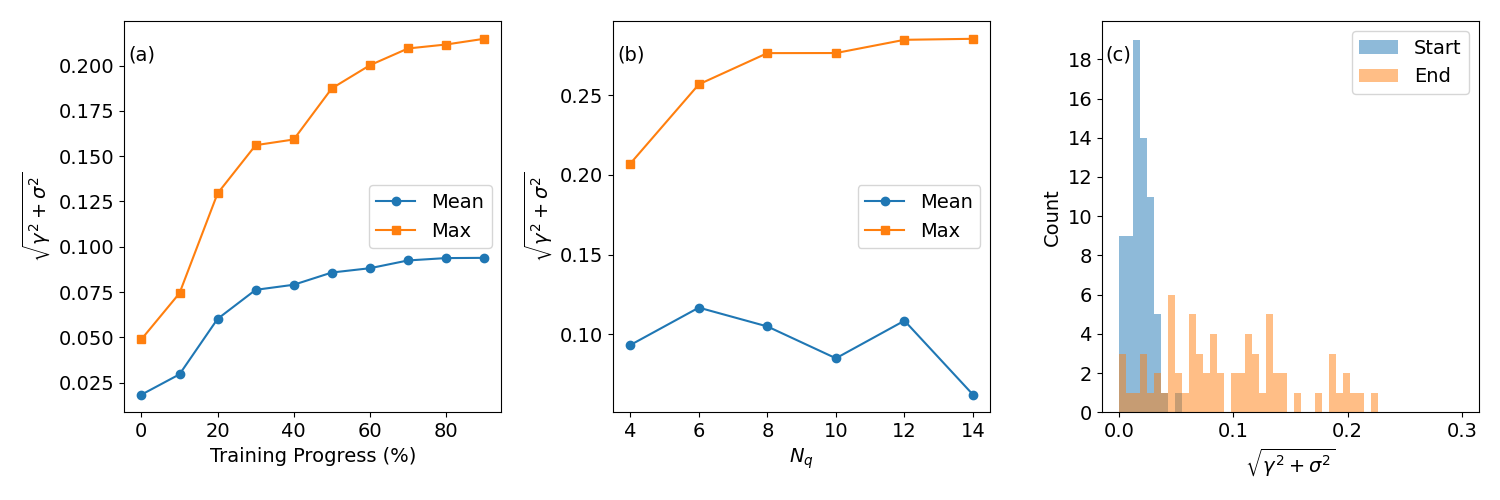}
    \caption{Analysis of how the $K_j\sim \sqrt{\gamma_j^2 +\sigma_j ^2}$ evolve as the training progresses. (a) This graph shows that the mean and maximum value of the distribution of $\sqrt{\gamma_j^2 +\sigma_j ^2}$ diverges as a 4 qubit model is trained. The mean and maximum are calculated with respect to 50 randomly chosen parameters. (b) The same quantities as a series of models on successively larger number of qubits is trained following the sub-net initialization strategy, again showing the divergence. (c) The distribution over $K_j$ values at the start and end of the 4 qubit training.}
    \label{fig:max_gradients_analysis}
\end{figure}

In Fig. \ref{fig:max_gradients_analysis}, we look at how the distribution of $\sqrt{\gamma_j^2 +\sigma_j ^2}$ changes as the optimization proceeds for a classification problem of the classes 0, 1, 2 and 3 of the MNIST dataset with the learning architecture described in Section \ref{sec:numerics}. Fig. \ref{fig:max_gradients_analysis}(a) shows the mean and maximum of the distribution for 4 qubits. We see that while both increase during the training, the difference between the two diverges with the maximum growing more quickly than the average. This can be seen also in Fig. \ref{fig:max_gradients_analysis}(c) which shows the distribution at the end of training is wider compared to that at the start. Fig. \ref{fig:max_gradients_analysis}(b) shows a similar phenomenon as the number of qubits in the model increases, in accordance with the sub-net training procedure in the next section.

The increasing value of the maximum implies that, for gradient-based methods, the learning rate schedule will need to be set so that the learning rate decreases as the training progresses. Physically, this corresponds to the parameter updates being smaller in magnitude so as not to miss the minimum. The exact coordinate scheme, in contrast, will not need to be constrained by the maximum value of $K_j$ but can instead make the maximum possible reduction of the loss function at each update.

While the above discussion focuses on comparing the exact coordinate update method vs gradient methods close to convergence, another advantage emerges far from convergence as well. From Eq. \ref{eq:loss_j}, we see that there exist saddle points in the optimization landscape which correspond to all $\theta_j\to\theta_j^*+\pi$. Gradient based methods, including optimizers like ADAM, will get stuck at such points as can be seen from Eq. \ref{eq:gd_update}. However, the exact coordinate update method will not get stuck and Eq. \ref{eq:ec_update} shows that it will reduce the loss function by the maximum allowed value of $2\sqrt{\gamma_j^2 +\sigma_j ^2}$. The number of such saddle points may grow exponentially with the dimension, thus making the exact coordinate update method more reliable even far away from convergence compared to gradient-based methods.

Finally, we see that the exact coordinate update method described here is guaranteed to decrease the cost function at each step until a local minimum is found, at least in the case that there is no stochasticity arising from the use of finite shots or finite batch size. Such a guarantee is not available for classical machine learning models regardless of which optimizer is used to train them. As we show in the numerical results section, practically, the convergence with this method works even when a fixed number of shots and random batches are used during training. 

In this sub-section, we have done an explicit comparative analysis for the exact coordinate update method compared to gradient methods. We expect similar outcomes to hold in comparing our method to other classical optimizers which depend upon analyzing the cost function in the neighborhood of the current parameter vector in order to make an update.

\section{Sub-net Initialization}\label{sec:subnet}

\begin{figure}
    \centering
    \includegraphics[width=\linewidth]{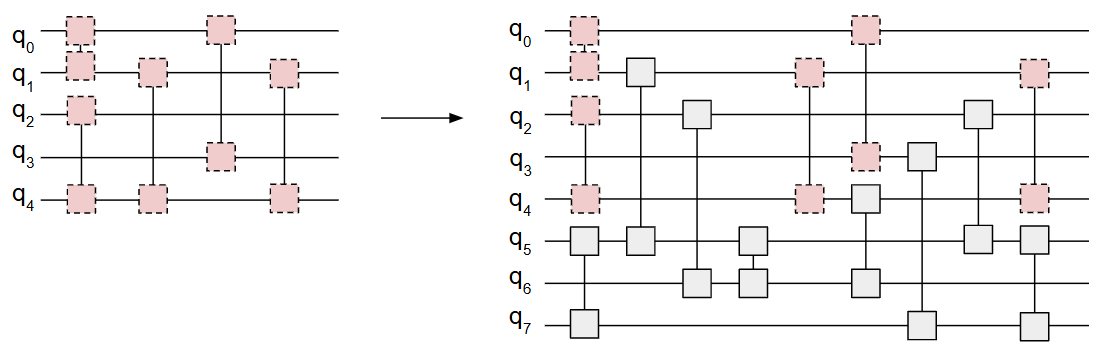}
    \caption{Example of an initialization from a sub-net. The model on the left has fewer parameters and acts on smaller input data. It is trained first and then used to initialize the corresponding sub-network in the model on the right (pink boxes, broken borders). Of the remaining parameterized nodes (gray boxes, solid borders), those connected to the sub-network are initialized as identity. The nodes not connected to the sub-network can be arbitrarily initialized. }
    \label{fig:subnet}
\end{figure}

We now turn to the problem of how to set the initial parameterization of quantum models so that they can avoid barren plateaus. This is a non-trivial issue which arises because the variance of the gradient for deep models goes to zero exponentially fast in the number of qubits. The issue is not confined to gradient based methods as even non-gradient based methods will encounter it if they rely on measuring the cost function in the $p$-dimensional neighborhood of a randomly initialized point\cite{Arrasmith2021effectofbarren}. Practically, this manifests as the training failing to reduce the cost function even many iterations after initialization.  

One set of approaches to mitigating barren plateaus involves restricting the ansatz structure, though these ansatzes often turn out to be susceptible to dequantization. Other approaches include specific training protocols which optimize a subset of parameters at a time such as layer-wise training \cite{Skolik2021}. Another approach involves initializing the parameters so that the ansatz consists of blocks of identity operators \cite{Grant2019initialization}. These techniques work with fixed input data and often with fixed model architectures. 

Here, in compatibility with the data encoding approach in Section \ref{sec:bit-bit}, we design a technique we call sub-net initialization which is depicted in Fig. \ref{fig:subnet}. The technique utilizes an incremental strategy to initialize large quantum models which take longer bit strings as input with parameters from smaller trained models which take more compact bit strings as input. It can be described as follows: let's visualize the quantum model as an `entanglement net' which consists of entangling nodes. We start the training on the dataset with an entanglement net that acts on only a small number of qubits and train it either for a given number of iterations or until the loss function stops changing. After this is completed, we construct a new model with more qubits. According to the bit-bit encoding scheme, we can now add more bits for each data sample while still respecting the data bit - to - qubit mapping of the previous model. We can also add more nodes to the entanglement net, however ensuring that the previous entanglement net is a sub-network of the current one. Under these conditions, the parameters of the smaller net can be used to initialize this sub-network and the value of the loss function remains the same. The rest of the nodes are initialized as identity if they are connected to the sub-network, or arbitrarily initialized otherwise. Training of the larger entanglement net now starts with updating the new nodes. The value of the loss function of the larger net at the start of its training is equal to the final value of the loss function at the end of the training of the sub-net.

Since parameters are effectively added in an incremental fashion, this ensures that the new network is not initialized in a barren plateau. Fig. \ref{fig:sub_net_chain} shows examples of training a large network via a sequence of smaller nets. Since smaller nets run faster, partial training on a series of expanding sub-nets also saves time compared to training from scratch on a large entanglement net. 

We note that an analog of sub-net initialization is not possible in classical deep learning models because their functional form does not allow for an exact transfer of parameters from smaller to larger nets. Sub-net initialization is similar to warm-start techniques that have been previously explored in variational quantum algorithms with the main difference being that here the entanglement net is free to grow as more quantum computational resources, in the form of qubits or gates, become available with the loss evolution always being restarted from the exact point where the previous training stopped.

\section{Numerical Results}\label{sec:numerics}
\subsection{Model Architecture}
\begin{figure}
    \centering
    \includegraphics[width=0.7\linewidth]{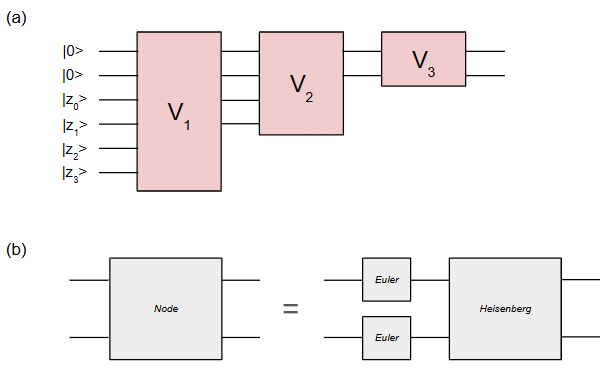}
    \caption{(a) The overall architecture of the quantum model used here has a cascading design where the number of qubits is reduced in successive layers. Each layer $V_i$ of the model consists of nodes that entangle pairs of qubits. At the output of each layer, some qubits are discarded, and others are passed to the next layer. (b) Each node consists of single-qubit `Euler' operators and a two-qubit `Heisenberg' operator. These are defined in Eq. \ref{eq:node_def}.}
    \label{fig:qcnn}
\end{figure}

We use a cascading architecture for the entanglement net similar to a quantum convolutional neural network \cite{Cong_2019}. It is depicted in Fig. \ref{fig:qcnn} (a). The architecture can be viewed as a succession of layers, with the number of qubits in layer $l+1$, $ N_{q,l+1}=2^{\lceil\log_2N_{q,l}\rceil-1}$, where the layer closest to the input is $l=0$. Except for the first layer, the number of qubits in each layer is thus a power of 2. The layer closest to the output has at least $N_y$ qubits, while the first layer has $N_q=N_x+N_y$ qubits. Within each layer, there is an entangling node between every pair of qubits, that is, they have all-to-all connectivity. High connectivity learning architectures have previously been shown to have advantages in expressivity \cite{daiwei_training} in quantum machine learning. However, they have mostly been avoided in numerical studies because of the hardness of training them from a random initialization \cite{PRXQuantum.3.010313}.

At initialization, $N_y$ qubits that will store the class at the output are set to 0 and the remaining $N_x$ qubits store the binary encoding of the data sample. At the output, only $N_y$ qubits are read out to determine the class.

Each entangling node acts on two qubits as shown in Fig. \ref{fig:qcnn}(b). It consists of single qubit Euler rotations acting on each qubit which correspond to unitaries in SU(2), followed by a `Heisenberg' unitary acting on both qubits. These are parameterized as follows.

\begin{align}\label{eq:node_def}
    U_{\text{Euler}}(\theta_1, \theta_2, \theta_3) = \text{Rx}(\theta_1) \text{Rz}(\theta_2) \text{Rx}(\theta_3)\nonumber\\
    U_{\text{Heisenberg}}(\theta_1, \theta_2, \theta_3) = \text{Rxx}(\theta_1) \text{Ryy}(\theta_2) \text{Rzz}(\theta_3)
\end{align}

where $\text{Rx}(\theta)$ is $\exp(-i\theta X)$ and $\text{Rxx}(\theta)$ is $\exp(-i\theta X\otimes X)$, and similarly for the other Pauli rotations. This construction is inspired by the general two-qubit gate decomposition in \cite{Vatan_2q}, and corresponds to a parameterized rotation of each qubit on its Bloch sphere before a parameterized entangling operation. An Euler unitary is also added to each qubit before it is measured.

Thus, the total number of parameters $p=9N_{\text{nodes}} + 3N_y$, where $N_{\text{nodes}}$ is the number of nodes in the entanglement net. The all-to-all connectivity results in models with a large number of parameters. Table \ref{tab:encoding_4_classes} shows an example of how the number of parameters scales as the number of qubits increases.

\subsection{Training Outcomes}
We test the above techniques on the MNIST dataset with all 10 classes and also a subset that consists of the digits $0, 1, 2, 3$. The training dataset consists of $80\%$ of the available data. The test dataset consists of 100 images per class chosen from the remaining images. For 4 classes, 2 qubits are reserved to store the class, and for 10 classes, 4 qubits are reserved. The rest of the qubits are used to load the binary encoded data at the beginning of the circuit. The circuits are run on the statevector simulator backend of the Qiskit software package \cite{qiskit}.

During training, the coordinates are updated sequentially one at a time, and we repeat the entire sequence 3 times for each model size. For each coordinate update, we use a random batch to calculate the values of the loss function that are input to Eq. \ref{eq:update}. We use heuristics to determine the batch size and number of shots as follows. For the batch size, we find that the loss function as measured on the test dataset is unstable if the batch size is too small. We therefore want the number of data samples in the random batch to be large enough so that the batch distribution is close enough to the test distribution. Therefore, before training begins, the batch size is determined by calculating the Kolmogorov-Smirnov (KS) statistic between data confined to the first PCA direction for a random batch and the test dataset. The random batch size is increased until the KS statistic meets a given threshold. Here, we use a batch size of 150. In principle, for each parameter update, it should suffice to use two measurements as shown in section \ref{sec:coord_update}. However, because of the stochasticity induced by the finite batch size, we find that the training is more stable if all three values of the loss function in Eq. \ref{eq:update} are calculated for the batch in each update. 

For the number of shots, at the start of training, we want the cost function to be determined precisely enough so that it can differentiate between the different classes. Therefore, we set the number of shots as the inverse squared of the minimum Wasserstein distance between all pairs of classes from the first PCA direction. This is rounded up to the nearest multiple of 100. For 4 classes, we start with 900 shots and for 10 classes, we start with 2300 shots. Further, at each repetition of the entire coordinate update sequence, we increase the number of shots by 1000.

\begin{figure}
    \centering
    \includegraphics[width=0.7\linewidth]{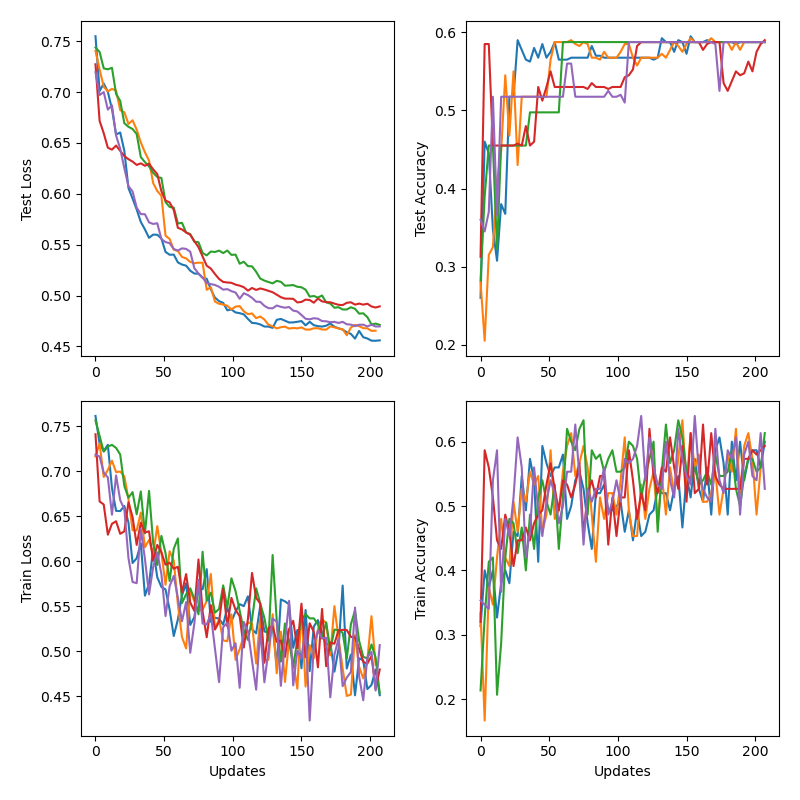}
    \caption{The evolution of loss and accuracy for training a 4 qubit model for 5 different training runs to classify the 0, 1, 2, and 3 classes of the MNIST dataset. The top row shows results for the test dataset which remains fixed throughout training and the bottom shows the results for the training batch which is randomly created for each update. The bit-bit encoding described in Section \ref{sec:bit-bit} is used to load and read out the data and exact coordinate update technique to evolve the parameters.}
    \label{fig:4_qubits_test_training}
\end{figure}

We first validate the framework for a small model. Fig \ref{fig:4_qubits_test_training} shows the results for training the 4 class dataset on 4 qubits for 5 different runs. The parameters are randomly initialized between $0$ and $0.4\pi$.  Each run corresponds to a different seed in the overall random number generator of the program. We show how the loss and accuracy for both the training and test dataset evolve as a function of the number of updates. While the training is noisy, the test loss converges relatively smoothly. We see that all of the trainings succeed although there is some spread in the final value of the loss. These models store only 2 data bits and can still train 4 classes up to $\sim 58\%$ accuracy.

\begin{figure}
    \centering
    \includegraphics[width=0.7\linewidth]{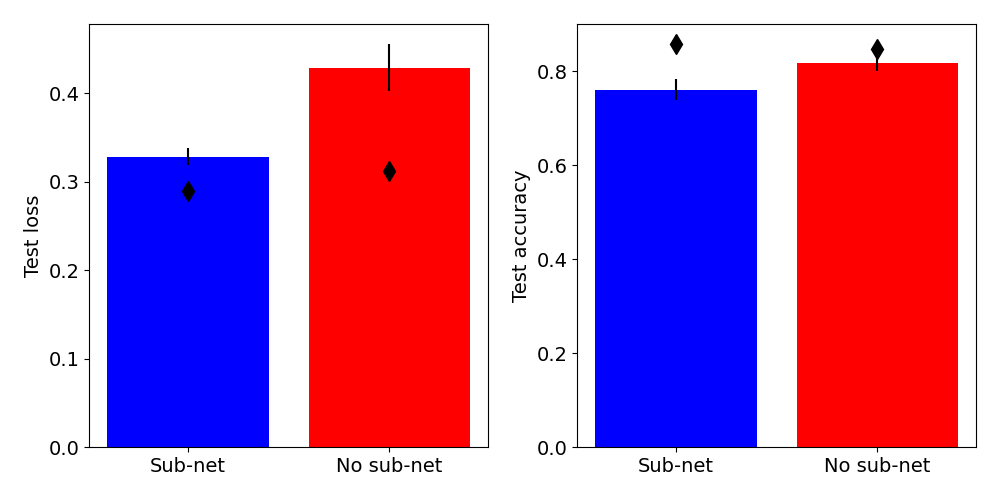}
    \caption{The bars compare mean test loss and accuracy for an 8 qubit model with and without sub-net training for the 4 class dataset. The diamonds depict the lowest loss and highest accuracies achieved in each instance. The error bars show the standard error calculated across 5 different training runs. The bit-bit encoding described in Section \ref{sec:bit-bit} is used to load and read out the data and exact coordinate update technique to evolve the parameters.}
    \label{fig:8q_sub_net_vs_no_sub_net}
\end{figure}

Next, we examine how the sub-net initialization method fares compared to training a larger model from scratch. We analyze this for an 8 qubit model for the 4 class dataset. In Figure \ref{fig:8q_sub_net_vs_no_sub_net}, the `sub-net' method trains a sequence of $4\to 6 \to 8$ models, and the `no sub-net' method directly starts training on the 8 qubit model with parameters randomly initialized between $0$ and $0.4\pi$. We do this with 5 independent runs. We see that the sub-net method results in lower average value of the loss as well as the highest accuracy model at $\sim 85\%$. However, note that the mean accuracy of the no sub-net method is somewhat higher. This indicates that intermediate values of loss do not perfectly reflect the accuracy, that is, decreasing loss does not necessarily imply increasing accuracy. Future work will explore varying the loss function so that there is a closer relationship between the two while maintaining the ability to update the parameter efficiently. 

\begin{figure}
    \centering
    \includegraphics[width=\linewidth]{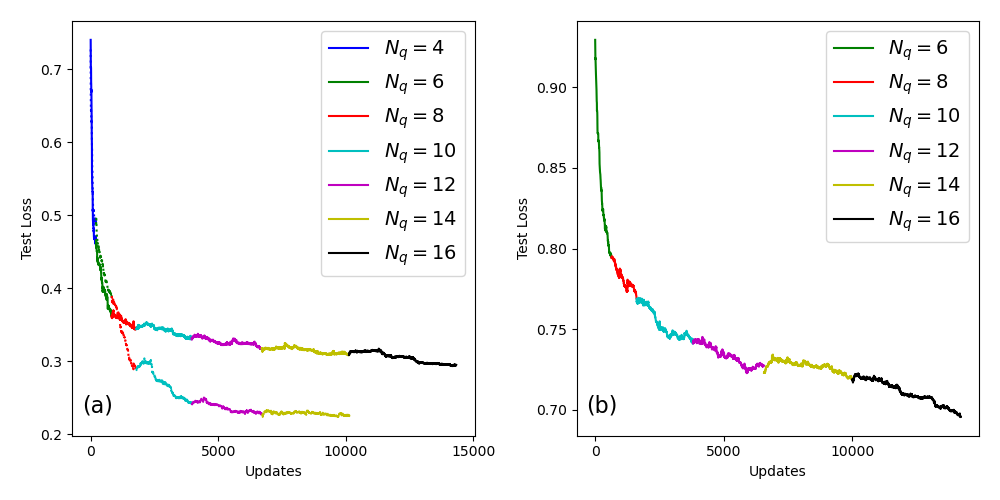}
    \caption{Test loss progress using bit-bit encoding, exact coordinate updates, and the sub-net initialization method for (a) two different training runs to classify the 4 classes - 0, 1, 2, 3 - of MNIST, and (b) one training run to classify all 10 classes of MNIST. Each color represents the performance of the model for a fixed number of qubits, with each size beginning where the previous size concludes.}
    \label{fig:sub_net_chain}
\end{figure}

Finally, Fig. \ref{fig:sub_net_chain} (a) and (b) show training up to 16 qubits for 4 classes and 10 classes respectively. Each color in the plot shows training for a different qubit size, and the training for each qubit size picks up from where the previous one ends. One such training run is shown for 10 classes and 2 are shown for 4 classes. In all cases, the loss decreases consistently. However, it is interesting to note that one of the 4 classes training runs results in a much lower loss function at 8 qubits than the other. This indicates that there are local minima in which the training can get stuck. The sub-net initialization strategy then allows a straightforward way to deal with such local minima. Namely, one can train a smaller model several times - picking the best performing one to move to the next stage. Since training a smaller model is less expensive and time-consuming than training a bigger one, this fail-fast strategy will win out in eliminating local minima compared to trying to escape from a local minima in a large model. As far as we know, such a strategy does not exist for classical deep learning, and is thus another advantage for training large quantum machine learning models. Future work will involve systematically using this method to push the performance of quantum machine learning models.

\section{Harnessing Near-term Quantum Computers to Build Large AI Models}
The framework presented here enables the utilization of near-term quantum computers for building large quantum models for machine learning or artificial intelligence applications. The bit-bit encoding scheme allows the expansion of the encoded bit string as the number of available qubits increases, the exact parameter update scheme ensures models can be trained until convergence, and the sub-net initialization strategy allows smaller models to be reused for the training of larger ones. Therefore, for example, if a 40 qubit quantum computer is available this year that can support a requisite number of entangling gates, the parameters of a model trained on it can be used to initialize a larger model the next year when a 50 qubit quantum computer is available, and the training of the larger model does not need to start from scratch. In this way, training on near-term quantum computers can be viewed as a step towards training models on larger computers, and is thus of concrete value. Even if a model that can be trained today does not reach the baseline performance required to beat a classical model, it remains a crucial step in training a larger model anticipated to outperform classical counterparts. To our knowledge, this is the first time such a clear argument has been presented for the direct utility of near-term quantum computers which are beyond the practical simulation capabilities of classical computers.

\section{Discussion and Future Directions}\label{sec:discussion}
In this work, we have proposed methods that address three critical issues in scaling quantum learning. First, we introduced a technique to load only the most predictive bits from a data sample into the quantum computer via a binary encoding technique that scales efficiently with the number of dimensions of the data samples. This is reminiscent of techniques from many-body physics where only the most correlated degrees of freedom are treated by expensive computational methods like exact diagonalization while the rest are treated approximately using mean-field techniques. Bit-bit encoding is very general and in fact, it is used in classical machine learning for large language models, though in that case the goal is not to compactly encode the data. Future work will involve improving upon this technique and also extending it to problems beyond classification such as object detection. 

Second, we outlined a training technique that allows for exactly updating parameters one at a time, thus guaranteeing convergence to a local minimum. An analogous technique does not exist for classical neural networks and may thus imply that quantum neural networks have comparatively better convergence properties at scale. We remark that optimizing parameters sequentially is typical of variational techniques in quantum many-body physics such as the density matrix renormalization group. We also note that the coordinate update technique can be applied to any variational quantum algorithm in which the parameterized unitary can be written as a sequence of independent Pauli rotations as in Eq. \ref{eq:U_exp_paulis} and is not restricted to quantum machine learning.

Third, we addressed the barren plateau problem through the sub-net initialization method. Again, the analogue of such a method which allows for exact transfer of parameters from smaller to larger networks does not exist for classical neural networks. This scheme does not require restrictions on expressivity or architecture that are common in other methods of avoiding barren plateaus.

In the numerical results section, we showed that just a few bits extracted efficiently from each data sample can be loaded onto a quantum computer for the purpose of classifying it, thus effectively addressing the data loading problem of quantum learning. Further, we demonstrated training very high-connectivity quantum models that are usually thought to be untrainable due to barren plateaus. This high connectivity is also leveraged in the `attention' mechanism in classical transformer architecture \cite{vaswani2023attentionneed}, and is likely essential to implementing quantum machine learning models that can encode many-body correlations that take a large amount of resources to implement classically, which in turn leads to unsustainable energy requirements for classical machine learning. Further, the advantages in training outlined in this paper, namely the guarantee of converging to a minimum, and the ability to escape from local minima early while training smaller models, may be key to understanding how to utilize quantum computers in machine learning.

A remaining challenge in training large quantum models is the lack of backpropagation scaling in the number of evaluations required to update all the parameters \cite{abbas2023quantumbackpropagationinformationreuse}. If every parameter has to be updated one at a time via the exact coordinate update method presented here, then the number of measurements is proportional to the number of parameters on average. Therefore, we need a smart way to pick which parameter to update at each step. We hypothesize that this can be done using the local graph structure of the quantum model and leave the exploration of this to future work.

A common theme in this work has been to leverage uniquely quantum properties of the learning framework. The universal approximation with respect to bit-bit encoding, exact coordinate updates via only two loss function measurements per sample, and the sub-net initialization are unique to quantum machine learning and do not have direct analogues in classical deep learning. Overall, we do not see compelling physical arguments for quantum machine learning to benefit from methods that work well for classical deep learning. We envision that future progress may also come from further developing an understanding of the physics of quantum models, rather than treating them as just another high-dimensional optimization problem.

Finally, while the emphasis of the numerical results in this paper has been validation of the techniques in simulation, there is a lot of scope for improving their performance and benchmarking how they perform on challenging datasets on quantum hardware. This will be the focus of future work.

\section{Software Framework}
The techniques in the paper are implemented in a software framework developed at Coherent Computing Inc. It can be made available upon request.

\newpage
\bibliography{references}

\end{document}